\documentclass[12pt,preprint]{aastex631}
\usepackage{natbib}
\usepackage{graphicx}
\usepackage{float}
\usepackage{latexsym}
\usepackage{longtable}
\usepackage{lineno}

\begin{document}

\title{The SMILES Mid-Infrared Survey}

\correspondingauthor{George Rieke}
\email{grieke@arizona.edu}

\author[0000-0003-2303-6519]{G. H. Rieke}
\affiliation{Steward Observatory, University of Arizona, Tucson, AZ 85721, USA, also Department of Planetary Sciences}

\author[0000-0002-8909-8782]{Stacey Alberts}
\affiliation{Steward Observatory, University of Arizona, Tucson, AZ 85721, USA}

\author[0000-0003-4702-7561]{Irene Shivaei}
\affiliation{Centro de Astrobiología (CAB), CSIC-INTA, Ctra. de Ajalvir km 4, Torrejón de Ardoz, E-28850, Madrid, Spain}

\author[0000-0002-6221-1829]{Jianwei Lyu}
\affiliation{Steward Observatory, University of Arizona, Tucson, AZ 85721, USA}

\author[0000-0001-9262-9997]{Christopher N. A.  Willmer}
\affiliation{Steward Observatory, University of Arizona, Tucson, AZ 85721, USA}

\author[0000-0003-4528-5639]{Pablo P\'erez-Gonz\'alez}
\affiliation{Centro de Astrobiología (CAB), CSIC-INTA, Ctra. de Ajalvir km 4, Torrejón de Ardoz, E-28850, Madrid, Spain}

\author[0000-0003-2919-7495]{Christina C. Williams}
\affiliation{NSF’s National Optical-Infrared Astronomy Research Laboratory, 950 North Cherry Avenue, Tucson, AZ 85719, USA}


\begin{abstract}

The Mid-Infrared Instrument (MIRI) for JWST  is supplied with a suite of imaging bandpass filters optimized for full spectral coverage in eight intermediate-width bands from 5 to 26$\mu$m and a narrower one at 11.3 $\mu$m. This contrasts with previous infrared space telescopes, which generally have provided only two
broad bands, one near 10 $\mu$m and the other near 20 $\mu$m. The expanded MIRI spectral capability provides new possibilities for detailed interpretation of survey results.  This is an important feature of the instrument, on top of its great increase in sensitivity and angular resolution over any previous mission. The Systematic Mid-infrared Instrument Legacy Extragalactic Survey (SMILES) was designed to take full advantage of this capability. This paper briefly describes the history of infrared surveys that paved the way for MIRI on JWST and for our approach to designng SMILES. It illustrates the use of the observations for a broad range of science programs, and concludes with a brief summary of the need for additional surveys with JWST/MIRI.

\end{abstract}


\section{A brief history of infrared surveys}

The deep mid-infrared sky (8 $\mu$m $<$ $\lambda$ $<$ 30 $\mu$m) is dominated by phenomena that are not prominent in the visible or near infrared. At the same time, the huge foregrounds emitted by the warm telescope and atmosphere partially blind observations from the ground and limit their sensitivity. Groundbased observations were critical in the formative years of infrared astronomy and still play a unique, essential role. However,  they by necessity focus on individual, compact sources.  An all-sky mid-infrared survey, or even one of a field of significant area, with the sensitivity to provide targets for follow up is not feasible from the ground because of the limitations imposed by the emission from the warm telescope. There is a great premium in surveys from space with cooled telescopes. 

The first all-sky mid-infrared survey is lost to history: the Hughes Celestial Mapping Program (CMP). A small telescope cooled by a closed cycle Viulleumier cooler and gimbaled from an Agena rocket was launched into a sun-synchronous polar orbit in 1971. Unfortunately, the cooler lines to the telescope started to leak shortly into the mission and only a few orbits of data were obtained. The results were presented  by the Hughes Aircraft Company as an attractive glossy black handout showing a projection of the sky with about 200 white dots superimposed, one for each source, and two numbers next to each dot. The handout also said ``secret'' in the upper right corner, but an enterprising Hughes  employee got rid of that and distributed the handout for publicity. We (George and Marcia Rieke) took the handout to a photographic plate measuring engine, determined the positions and identifications for a subset of the white dots, and deduced that the numbers were the flux densities at 10 and 20 microns. This let us construct an all-sky mid-infrared catalog. When we requested permission to publish this quasi-secret information, it was denied on the basis that if the Russians discovered the wavelengths, they might jam them in warfare (!?).  

The military had a strong interest in mapping the sky in the infrared, from a concern that systems designed to track warheads and military vehicles of any kind might instead lock onto some celestial object. There were a large number of military-sponsored surveys;  comprehensive descriptions are provided by Steven Price \citep{price1988,price2008}, who was a central figure in many of them. The results for many never made their way out of the military sphere and into the astronomical community. Foremost among those that did emerge was the HISTAR program, based on Gregorian telescopes with 16.5 cm (i.e., 6.5 inch) diameter primary mirrors, cooled with super-critical helium, and launched on sounding rockets from 1971 through 1974. HISTAR mapped in three bands, centered at 4, 11, and 20 $\mu$m. The relative band widths $\Delta W = \Delta \lambda/\lambda_0$, where $\Delta \lambda$ is the FWHM of the spectral band and $\lambda_0$ is its center wavelength, are respectively for the HI STAR bands 0.5, 0.55, and 0.4. For a southern extension of the catalog, a band at 27 $\mu$m was substituted for the one at 4 $\mu$m. The final catalog from this program included nearly 3000 sources. Some of the detections were difficult to confirm, either because the sources were extended, or because they were spurious associated with dust and debris launched along with the rocket. Nonetheless, this was the first astronomically revolutionary mid-infrared all sky survey. A parallel effort was funded by the Air Force Cambridge Research Laboratory (AFCRL) from the ground under the supervision of Frank Low, but it only detected a single source independently of the rocket results, AFGL 490. This provides a dramatic illustration of the challenges in surveying from the ground.  

The next revolution in mid-infrared surveys was with the Infrared Astronomy Satellite (IRAS) \citep{neugebauer1984}, launched in 1983. IRAS mapped nearly all of the sky in mid-infrared bands at 12 and 25 $\mu$m, with $\Delta W$ of 0.54 and 0.44, respectively (IRAS also included far-infrared bands at 60 and 100 $\mu$m). Although the telescope was of 60 cm aperture and capable of 10$''$ resolution in its mid-IR bands, the resolution was limited by the need to use relatively large detector fields of view for mapping efficiency: $0'.75 \times 4'.5$ arcmin. For the first time, IRAS provided measurements to a level that made infrared observations a common tool across all of observational astronomy. In 1996, the Midcourse Space Experiment (MSX) was launched and surveyed the Galactic Plane and selected areas with the SPIRIT III telescope, in four bands between 8 and 22 $\mu$m.  Another forward leap was the Widefield Infrared Survey Explorer (WISE) all-sky survey \citep{wright2010},  conducted at two shorter wavelength bands plus ones at 11.56 and 22.09 $\mu$m, $\Delta W$ = 0.48 and 0.19 respectively, and which extended the reach of the infrared even further than IRAS. The Akari mission \citep{murakami2007} also conducted an all-sky survey at 9 and 18 $\mu$m with $\Delta W$ $\sim$  0.9 and 0.8 respectively. 

The first cooled space telescope designed for pointed observations of individual sources was the Infrared Space Observatory (ISO).  Dedicated areal surveys have been conducted with the mid-IR camera on ISO   at 6.7 and 15 $\mu$m, $\Delta W$ = 0.52 and 0.40, respectively \citep{rowan2004}. Akari had a more complete complement of filters that were used to map the ecliptic poles \citep{kim2012}. Spitzer, with its large arrays, good pointing, and good agility was a mapping machine covering significant areas at 5.6, 7.8, and 24 $\mu$m, $\Delta W$ = 0.25, 0.37, and 0.22 respectively \citep[e.g.,][]{huang2004,papovich2004}; small areas were also surveyed at 16 $\mu$m using the IRS peakup array in the GOODS fields \citep{teplitz2011}. 

From this summary, prior to JWST our knowledge of the mid-infrared sky was largely based on observations in two spectral bands, one near 10 $\mu$m and the other near 20 $\mu$m. The bands were generally so broad in spectral coverage as to make derivation of accurate photometry in physical units challenging. That is, the bandpass corrections to allow for differing source spectra across the spectral band were potentially large and hence uncertain. Nonetheless, these efforts have covered large areas on the sky, providing the preparation needed for efficient use of JWST. Spitzer was particularly important in this regard, since its agility and relatively large detector arrays allowed it to map at high efficiency, and the duration of the mission allowed for mid-IR surveys of significant depth. These surveys were the most sensitive to embedded star formation out to z $\sim$ 2.5 \citep{perez2005, elbaz2011} and were used in multiple ways to search for and study AGNs, as examples. They covered many square degrees, supporting studies of the infrared properties of galaxies out to cosmic noon as a function of environment as well as supplying large samples of infrared-emitting galaxies for study in other ways. 

To build on this legacy, the Mid-Infrared Instrument (MIRI) on JWST  provides the ability for selected small-area surveys of unprecedented sensitivity and angular resolution. It also provides far more spectral information than any previous capability, with eight spectral bands between 5 and 26 $\mu$m, each with $\Delta W$ $\sim$ 0.2, and contiguous in wavelength  so the ensemble acts as a kind of very low resolution spectrometer. The Systematic Mid-infrared Instrument Legacy Extragalactic Survey (SMILES, PID 1207) was designed to take full advantage of all three gains. Through Cycle 3, only one other MIRI survey combines most  bands - that of the CEERS program \citep{yang2023}  -- and it covers a significantly smaller area with less ancillary data. SMILES is therefore a somewhat unique resource until larger community surveys are approved. We describe its rationale and planning in this paper. The technical design, data reduction and photometric catalog, and data release products are presented in Alberts et al. (2024).

\begin{figure*}
\epsscale{0.95}
\plotone{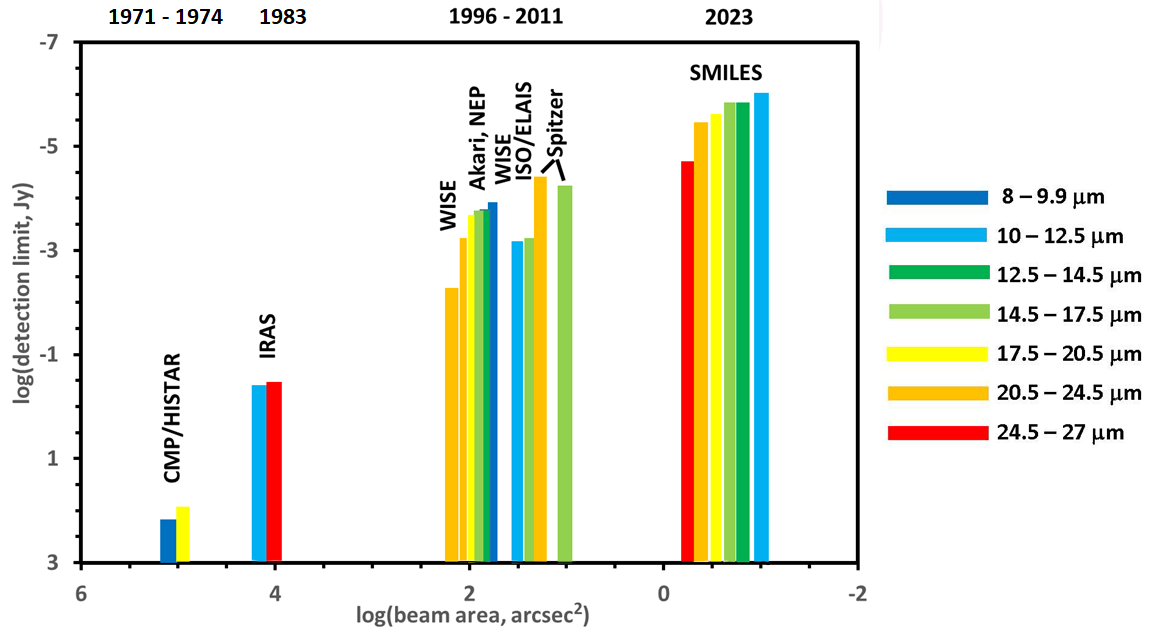}
\caption{A summary of the characteristics of mid-infrared surveys operating over the range 8 $< \lambda <$ 30 $\mu$m. The detection limits and beam ares are plotted in reverse order so that better is always to the upper right. The bands in a survey are color coded, from dark blue for for 8 - 9.5 $\mu$m to red for 24.5 - 27 $\mu$m. This figure illustrates the nearly unique status of SMILES in terms of sensitivity, angular resolution, and number of bands between 8 and 26 $\mu$m (the CEERS MIRI survey \citep{yang2023} includes all the bands shown but one and covers  $\sim$ 20\% as large a field). To avoid clutter, the MSX/SPIRIT III survey has been left out of figure; the sensitivity values are similar to those from IRAS and the beam area falls modestly to the left (larger values) compared with the other 1996-2011 surveys. }
\label{surveys}
\end{figure*}

Figure~\ref{surveys} is a pictorial summary of the discussion in this section. In fairness it needs to be pointed out that the CMP/HISTAR, IRAS, and WISE entries are for all-sky surveys, while the others are for selected areas: (1) Akari, north ecliptic pole \citep{kim2012}; (2) ISO/ELAIS \citep{rowan2004}; (3) Spitzer 16 $\mu$m \citep{teplitz2011}; and (4) Spitzer 24 $\mu$m \citep[e.g.,][]{dole2004, scott2010} largely for the GOODS fields. The figure covers 50 years, starting from the CMP in the early 1970's.  From then to SMILES and  CEERS,  it shows a growth in sensitivity by 7 - 8 orders of magnitude, in angular resolution by 6 orders of magnitude, and in spectral bands by about a factor of four. 

\section{The design of SMILES}

Identifying an unbiased sample of AGNs has been a long quest in astronomy \citep[e.g.,][]{keel1980, huchra1982}.  The contributions to this search made with Spitzer data emphasized how a panchromatic approach is essential \citep{cardamone2008, donley2008, delvecchio2017}. The combination of Spitzer's infrared capabilities and hard X-ray observations revealed a large number of heavily dust-obscured AGN not known previously \citep[e.g.,][]{lacy2004,donley2005,stern2005,alonso2006,donley2008, delmoro2016}. 

The original impetus for SMILES was to greatly expand our  understanding of these objects.  Key aspects of this population remained undetermined, including their number density and fraction relative to all AGN, with estimates ranging from $10 ~\rm{to} >50\%$ \citep{mendez2013, delmoro2016, ananna2019} of the total population. The Spitzer searches were efficient in identifying luminous AGN with roughly power-law emission signatures in the IRAC bands up to $z\sim3.5$ \citep{donley2012}, i.e., relatively lightly obscured Type 1 AGNs. Incorporation of the MIPS 24$\mu$m band could extend the baseline for identifying obscured AGN, but contamination by emission from star formation required careful SED fitting to disentangle the relative contributions of emission by stellar-heated dust and that from an AGN \citep[e.g.,][]{lyu2022}. That is, the Spitzer searches for the obscured population were limited by the lack of detailed spectral coverage between the longest wavelength IRAC band at 8 $\mu$m and the MIPS 24 $\mu$m band.   These limitations are removed for MIRI, which has the sensitive, continuous coverage from 5 - 25.5 $\mu$m required to (1) probe at typical (i.e., z $\lesssim$ 4) AGN redshifts the rest-3 - 5 $\mu$m wavelength regime where the stellar emission is at a minimum  and even moderately low-luminosity AGNs can be identified above it; and (2) cover rest-6 - 9 $\mu$m, where PAH bands indicate whether the infrared emission is predominantly powered by young stars.  

A full treatment of the obscured AGN population up to cosmic noon therefore motivates a deep survey with MIRI, which can provide  continuous coverage at a spectral resolution of $\sim$ 20\% from 5.6 through 25.5 $\mu$m.  Extrapolation from the Spitzer studies \citep[e.g.,][]{donley2008} led to the conclusion that an area $\gtrsim$ 30 square arcmin would need to be surveyed to obtain adequate statistics. To extend as far in redshift as possible, the survey needed to include deep observations in the 21 $\mu$m band. The time available in the GTO allocation then set an integration time of $\sim$ 2000 seconds in F2100W. Integration times were set for the  shorter wavelength bands on the basis of a fiducial obscured-AGN SED that is fairly flat from 12 to 21 $\mu$m and then falls steeply (as $\sim$ $\nu^{-2}$) toward shorter wavelengths. As presented in \citet{alberts2020}, this strategy was expected to allow detection of a Mrk 231-like SED at $1 (10)\%$ of Mrk 231's bolometric luminosity, e.g. $10^{10}$ ($10^{11}$) L$_\odot$  at $z\sim1$ ($z\sim2$). The 25.5 $\mu$m band was included with a shorter integration; the high backgrounds make it impossible to match the detection limits in the other bands with reasonable exposure times. However, simulations indicated that this band would be useful in identifying obscured AGN out to z $\sim$ 1, in analyzing relatively bright AGNs, and for stacking analyses. The pre-launch rationale and expectations for this survey are summarized in \citet{alberts2020}. In fact, the survey is significantly more comprehensive than anticipated because of the improved  sensitivity of MIRI compared with prelaunch predictions (Alberts et al 2024, in prep) as summarized in Table~\ref{tab:inttimes}. 

\begin{figure}
\epsscale{0.85}
\plotone{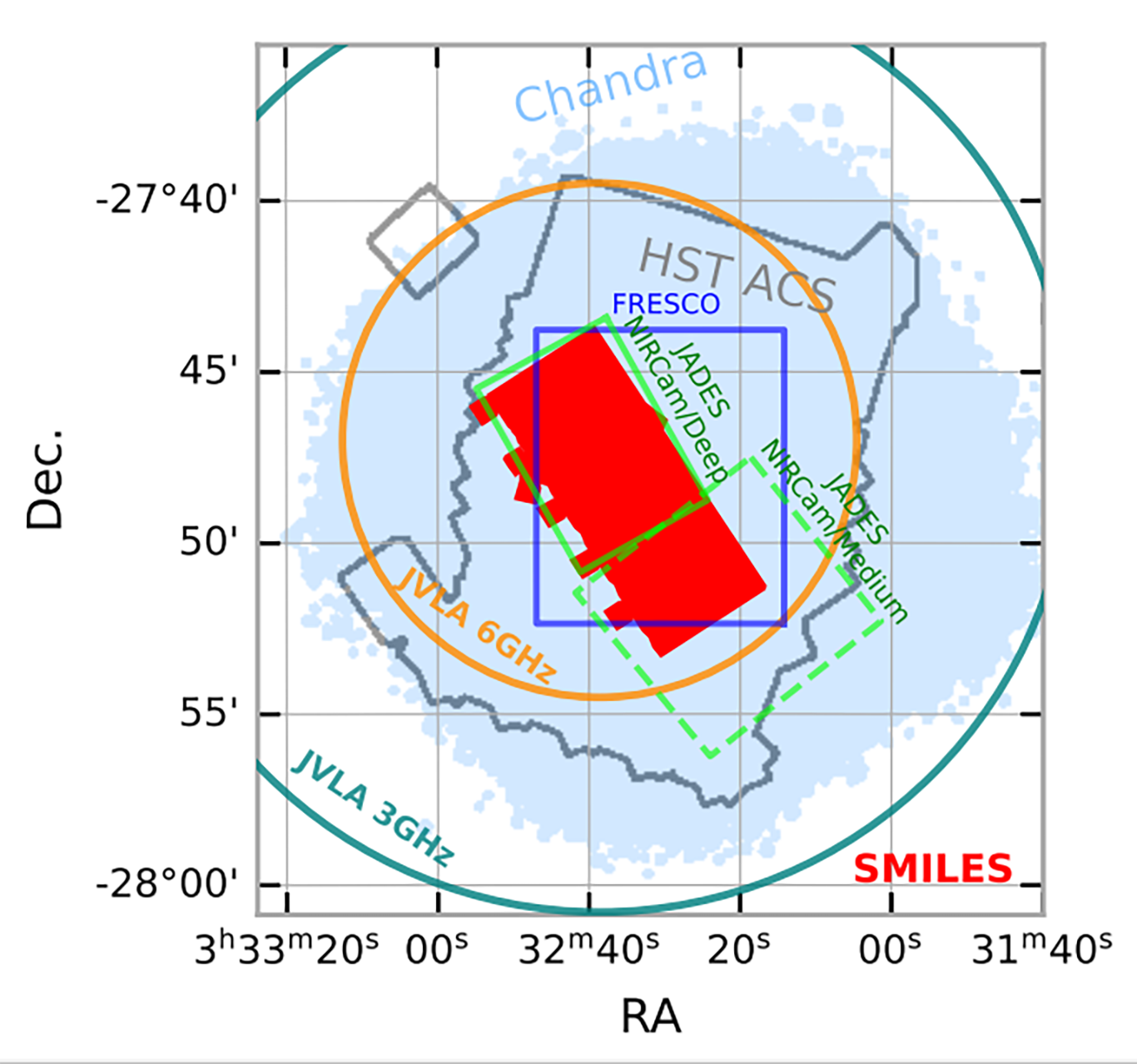}
\caption{SMILES footprint (red) relative to other HUDF surveys:  JADES NIRCam deep (solid green line) and medium (dashed green line), 
FRESCO NIRCam/grism coverage (solid blue line), HST ACS GOODS-S coverage (thick gray line), {\it Chandra}
    X-ray coverage (light blue shaded region) and JVLA radio observations at 3 GHz (dark green lines) and 6 GHz (dark orange lines). }
\label{layout}
\end{figure}

To put obscured AGN in the context of the full AGN population, the GOODS-S field \citep{giavalisco2004} was chosen for its extensive multi-wavelength coverage, including ultra-deep coverage in the X-ray with Chandra, as well as NIRCam and NIRSpec under the JADES program \citep[e.g.,][]{rieke2023}, optical/UV imaging with CANDELS \citep{grogin2011}, and multiple spectroscopic programs \citep[e.g.,][]{momcheva2016, bacon2017}. For use specifically to enhance the SMILES-based science, we included three pointings of dedicated NIRSpec MSA observations. The SMILES footprint relative to the ancillary data in GOODS-S is presented in Figure~\ref{layout}.  

The basic SMILES survey design was found to enable an entirely different set of science investigations. The detection levels at 21 $\mu$m are  fainter than the confusion limit for Spitzer MIPS at 24 $\mu$m by a factor of nearly ten \citep{dole2004, rodighiero2006}. This allows measurement of the obscured star formation equivalent to 10 M$_\odot$  yr$^{-1}$ or LIR$\,\sim10^{11}\,L_\odot$ for a star-forming galaxy at $z\sim2$. This capability can probe a few thousand galaxies with SFRs down to well below the main sequence at this redshift for galaxies with mass $\gtrsim$ 10$^9$ M$_\odot$ \citep{popesso2023}. To enhance this area of science,  the exposure times were increased by factors of 1.7 $-$ 3.5 at $10-18\,\mu$m from those required based on an AGN SED. The improved detection limits support studies of the morphology of the obscured star formation. They also enable probing the general behavior of the aromatic features at these redshifts.  The continuous wavelength coverage of MIRI greatly expands our previous capabilities in quantifying the nature of this small grain dust and the obscured star formation by fully sampling the mid-infrared regime (with spectral resolution $\sim$ 20\%).    

\section{Example projects}

\subsection{Proof of concept}

We now show that the original goals embodied in the design of SMILES have been successfully attained, based on JWST observations.

\subsubsection{Obscured AGNs}

\begin{figure*}
\epsscale{0.95}
\plotone{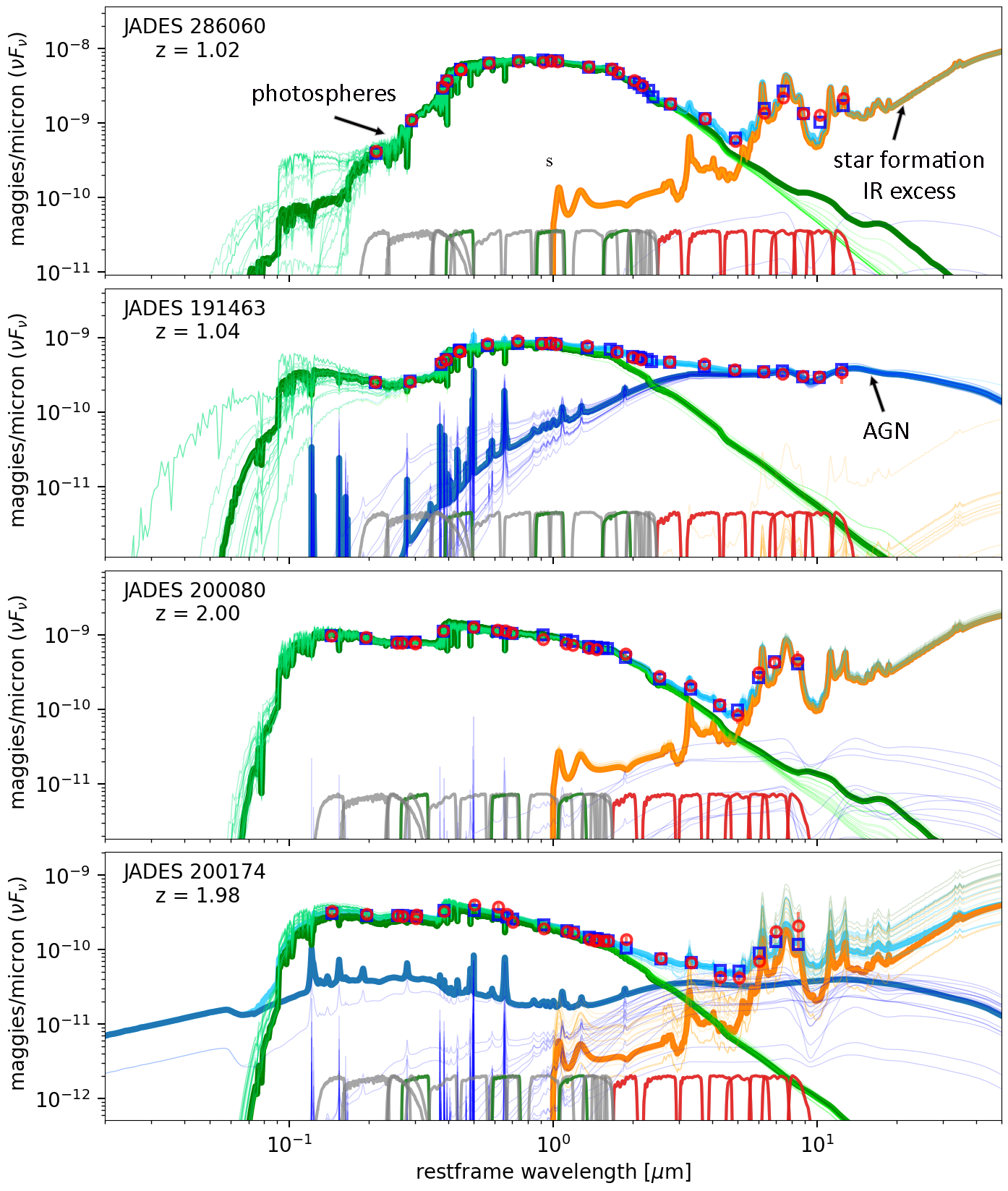}
\caption{Identification of embedded AGNs with SMILES data: the top and third frames show purely star forming galaxies at redshifts of $\sim$ 1 and 2. 
The second frame is a powerful obscured AGN at z $\sim$ 1, and the bottom frame is a mixed AGN/star forming galaxy at z $\sim$ 2, from \citet{lyu2023}. All of the filter profiles 
are shown along the bottom of the frames and the wavelengths have been shifted to the rest frame; in addition to SMILES data, the shorter wavelength measurements are from CANDELS \citep{grogin2011}, JADES \citep{rieke2023} and JEMS \citep{williams2023b}.  A model has been fitted to the photometry using Prospector: the green lines 
are for the stellar component, both the direct photospheric output and the infrared excess; the blue line is a template AGN SED. The measurements are shown as open circles with error bars, and the
blue squares are the result of synthetic photometry on the model. Very light lines show a range of SEDs for the AGN and star-forming templates. }
\label{agns}
\end{figure*}

The rationale for the SMILES program based on obscured AGNs was explained in the discussion of survey design. Figure~\ref{agns}, after \citet{lyu2023}, shows how the capability to identify PAH emission and other characteristics of star forming infrared excesses (i.e., the low level of dust warm enough to emit strongly at wavelengths $<$ 6$\mu$m) has allowed isolation of galaxies where the mid-infrared has a significant contribution from  the emission powered by AGNs. This can be compared with a previous paper, \citet{lyu2022}, where similar work was carried out using the Spitzer IRAC and MIPS bands, i.e., with no coverage between 8 and 24 $\mu$m. The reliability of the AGN identifications is greatly increased by the multiple bands, including finding that some of the cases thought to be AGNs in the earlier paper are dominated by star formation.  Altogether, \citet{lyu2023} find 111 AGNs in massive galaxies ($M* > 10^{9.5}$ M$_\odot$), a similar number of AGN candidates in lower-mass hosts, and about two dozen candidates at z $>$ 4,  all within the SMILES 34 arcmin$^{2}$ field. Fully 34\% of the AGNs in the massive galaxies were not known previously and reveal the previously suspected but undetected population of heavily obscured AGNs.  With this more complete census of AGN in hand \citep{yang2023,lyu2023}, we can now begin to explore the galaxy-AGN connection. In addition, AGN in dwarf galaxies at z $\sim$ 1 - 2 and at very high-z are poorly-charted in the past, and SMILES demonstrates the power of MIRI to contribute to these areas.

\subsubsection{Aromatic Bands}

\begin{figure*}
\epsscale{0.85}
\plotone{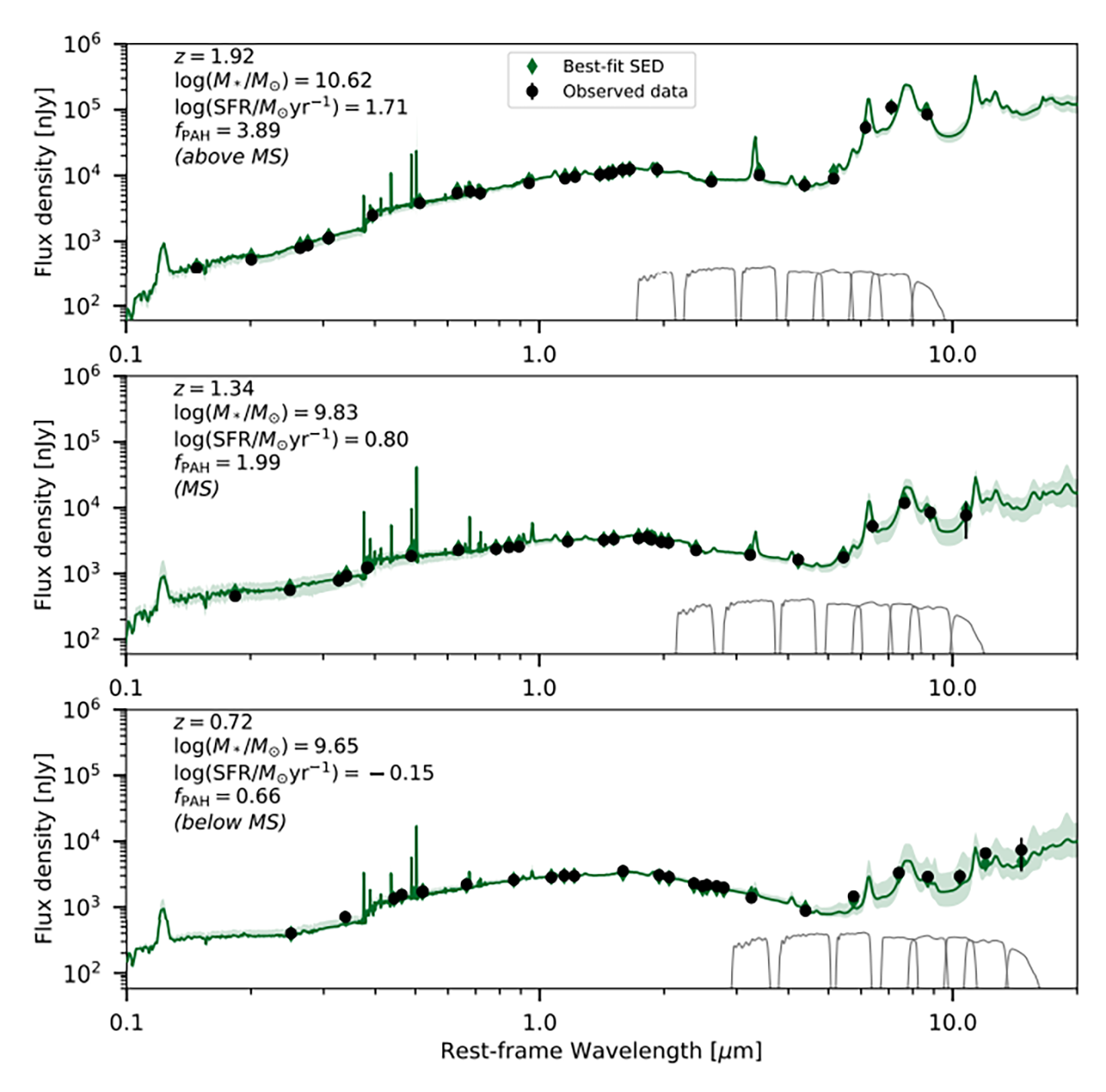}
\caption{Measurement of PAH band strengths with the SMILES photometry, for galaxies from z = 0.72 (bottom) to 1.92 (top), from \citep{shivaei2024} (all measurements shifted to the rest frame).
 The shorter wavelength bands that extensively trace the outout of the stars in the galaxies, are from the CANDELS \citep{grogin2011}, JADES \citep{rieke2023} and JEMS \citep{williams2023b} programs. The MIRI filter bands are shown 
 along the bottom of each frame with the corresponding fluxes as dots against a template-derived PAH spectrum. }
\label{pahs}
\end{figure*}

In star-forming galaxies, the mid-infrared spectrum is dominated by broad emission features arising from polycyclic aromatic hydrocarbons (PAHs), a type of small-grain dust. The behavior of the PAH bands with metallicity and environment is broadly studied and provides clues to their nature \citep[e.g.,][]{lai2020}. In addition, they are useful indicators of star formation rates in galaxies \citep{shipley2016}, an aspect that becomes critical for MIRI at high redshifts when SFRs must be determined from PAH bands shifted into the longer wavelength MIRI filters.  Figure~\ref{pahs} from \citet{shivaei2024} shows that fully sampling over all of the MIRI bands does indeed allow identification of the PAH bands and estimation of their strengths. That paper demonstrates that the PAH behavior with metallicity is very similar at z $\sim 1 - 2$ to that locally.  Judging from the scatter, the PAH strength determinations may be about as accurate as achieved locally using IRAC data \citep{marble2010}. Star formation rates for cosmic noon galaxies detected in SMILES determined via SED fitting \citep{shivaei2024} and through comparisons to gold standard Pa$\alpha$-based SFRs (Alberts et al, 2024b, in prep) have found that our initial estimates were conservative; we are able to robustly measure SFRs to well below 10 M$_\odot$/yr up to $z\sim2$ and down to masses of $\sim10^9$ M$_\odot$ \citep{shivaei2024}, probing the full main sequence and a factor of 30 below the MIPS confusion limit at the same redshift.  This is a significant advance over previous studies with Spitzer at these redshifts; for the first time, we can study the evolution of obscured luminosity and the fraction of dust in PAHs from $z\sim 0$ to 2 in the main-sequence galaxies down to stellar masses of $\sim 10^{9}\,M_{\odot}$ \citep{shivaei2024}.

The study also indicates that the PAH behavior is universal: the fraction of dust mass in PAHs, q$_{PAH}$,  is constant at $\sim$ 3.4\% above a gas metallicity of Z $\sim$ 0.5 Z$_\odot$ and decreases to $<$ 1\% at metallicities $\le$ 0.3 Z$_\odot$. That is, gas  metallicity traces the ISM conditions governing the formation and destruction of PAHs.

\subsection{Other investigations}

Without trying to be comprehensive, here we mention a few programs that also benefit from the SMILES design strategy. Florian et al. (2024, in preparation) study the  extent of the embedded star formation in high redshift (z $\sim$ 1 - 2) luminous infrared galaxies (LIRGs) using spatially resolved coverage of the PAH dust emission features.  
Previous work \citep[e.g.,][]{rujo2011} had suggested that LIRGs at these redshifts had more extended and diffuse star formation than local ones, leading to less-obscured SEDs. A consequence would be that, while local LIRGs often require a major interaction to cause matter to drift into their nuclei and ignite high levels of star formation \citep[e.g.,][]{sanders1996}, the process may be different at high redshift, possibly favoring mergers with a low mass galaxy or steady accretion of material. To maintain the highest possible angular resolution on the dust emission, Florian et al. focused on the 6.3 $\mu$m PAH feature and selected the galaxy image in the MIRI band where this feature lay, given the galaxy redshift. They find that, indeed, the high redshift LIRGs tend to be more extended than the local ones. In support of this behavior (indicating a different triggering mechanism for the star forming activity), the high redshift galaxies do not seem to be strongly disturbed at a statistically significant level (whereas local luminous LIRGs do tend to be clearly disturbed). The redshift distribution of the galaxies suitable for this study would be substantially reduced without the multiple MIRI bands in SMILES, and it is not clear that a statistically significant result would have been reached. 

Another example, which was not anticipated in the design of SMILES, is the study of the so-called little red dots (LRDs; \citet{matthee2024}), in which our moderate-depth MIRI imaging has played a large role 
\citep{williams2023a,perez2024}.  Previous measurements of these sources in the NIRCam bands up to 4.44 $\mu$m show a steadily rising SED, giving rise to speculation that they are dominated by obscured AGNs. 
This rise is continued through 7.7 $\mu$m with SMILES data, but starts to roll over at 10 $\mu$m. \citet{williams2023a,barro2024, perez2024} use the measurements in the 12.8, 15, and 18 $\mu$m bands to document the flattening more thoroughly (too few of the LRDs are detected well to SMILES depths at 21 $\mu$m). The shape of the SEDs beyond 10 $\mu$m (the rest near infrared) is important because it appears to be inconsistent with the pure obscured AGN model and
indicates that the SEDs are dominated by stars in the rest near infrared.  Even with deeper 21 $\mu$m data but skipping the intermediate bands, the result would be much more ambiguous than with the SMILES data. 

\subsection{Optimizing MIRI surveys}

The strategy adopted for SMILES is very different from the approved MIRI survey programs PIDs 1837, 5407, and 5893, which follow the traditional approach of two survey bands, one near 10 $\mu$m and the other near 20 $\mu$m. Although useful for discovery of AGNs and estimation of SFRs, the lack of all the intermediate MIRI bands loses a substantial amount of science as discussed above. As an example, for the integration times in our adjusted program (Table 1), the totals are divided equally between 10 and 21 $\mu$m together and the three intermediate bands together. In other words, doing all the bands resulted in reducing the integration at 10 and 21 $\mu$m by a factor of two for the same survey time. Equivalently, SMILES went 70\% as deep at 10 and 21 $\mu$m, which from the number counts at these wavelengths resulted in the detection of 20 - 25\% fewer sources at 10 and 21 $\mu$m (Stone et al. 2024, submittd to ApJ) than if the intermediate bands had been skipped. The alternative, which has been adopted by the other surveys, is to survey twice the area to similar depths.  Including the intermediate bands is a good trade for the immediate science return and certainly improves the legacy value of the data. The situation is slightly but not significantly different for surveying just at 10 and 18 $\mu$m such as in  PID 1837. 

A more challenging question is posed by PID 3794, which surveys at 10, 15, and 21 $\mu$m; a single intermediate band should certainly recover some of the science. To evaluate this strategy in more detail, we assume a source with the same flux density (in $\mu$Jy) at all three intermediate bands and compare integration times distributed as in Table 1 but with the results combined into a single weighted average vs.  putting the same integration time just into the 15 $\mu$m band. The two approaches both give measurements nominally at 15 $\mu$m and we find them to have virtually identical signal to noise. Obviously this result would change to some degree depending on the actual SED of a source, but it illustrates that the full set of MIRI bands can be obtained at very little extra cost compared with obtaining a subset of them. 


\section{Future directions}

JWST scheduling has put a huge emphasis on blind mapping with NIRCam; to date, there are $>$ 5100 arcmin$^2$ mapped in four or more filters and $>$ 3200 arcmin$^2$  mapped in six filters or more. In comparison, there are $\sim$ 113 arcmin$^2$ mapped with MIRI in four or more filters (at 7.7 $\mu$m and beyond)  and 46 arcmin$^2$ with six or more, i.e., about 2\% as much (there are larger-area MIRI surveys but only in two filters).  There is a range of important phenomena that can best be observed in the mid-infrared, e.g., dust heated by obscured star formation, the emission of obscured AGNs plus phenomena in the rest near infrared red-shifted into the mid-infrared, like the torus emission of AGNs, the redshifted near infrared and optical stellar photospheric emission (to estimate star formation histories and masses), and the detection of very high redshift (z $>$ 7)  galaxies including strong emission lines in specific MIRI filter bands.  
Our experience with SMILES has shown that it has two very important features: (1) its multiband design; and (2) the ancillary data in the HUDF where it is located. Surveys of similar characteristics would be very valuable in other fields where there is a strong set of ancillary data covering the full electromagnetic spectrum, such as the Hubble Deep Field, Groth Strip, and selected areas within the COSMOS field and the North Ecliptic Pole. In some cases, it may be advantageous to take advantage of existing surveys and focus on filling in the missing filter bands, in others suitable new surveys are needed {\it ab initio}.  Only with this broader set of surveys can we fully address issues such as cosmic variance and the characteristics of galaxies in large overdensities in the early Universe.

In parallel to this effort, much deeper surveys over small areas are needed to explore a different class of phenomena, such as the nature of optically dark galaxies such as little red dots, the fraction of obscured star formation in dwarf galaxies around cosmic noon,  and the presence of heavily obscured AGNs at redshifts $>$ 4. 
The needs for increased area and a mixture of shallow and deep surveys should be very familiar, since similar arguments have guided virtually all survey strategies at wavelengths besides the mid-infrared.

The mid-infrared (MIRI wavelengths) is arguably where JWST has the most unique contributions to make; there are no active plans for future missions there, and the gain over previous ones (e.g., Spitzer) is immense, a factor of 50 $-$100 in sensitivity, 50 in spatial resolution, and much expanded instrumental capabilities, as illustrated in Figure~\ref{surveys}.  JWST needs to leave a rich legacy in the mid-infrared for the future of astronomy, but to do so will require substantial increases in the area and spectral coverage of MIRI surveys. As shown by the broad application of data from SMILES, only by surveying in the full suite of MIRI bands will we leave a legacy with the flexibility to address future science objectives.

\begin{deluxetable}{lcccc}
\tabletypesize{\footnotesize}
\tablecaption{Integration times and final sensitivities\tablenotemark{a}}
\tablewidth{0pt}
\tablehead{
\colhead {band} &
\colhead {original} &
\colhead {5 $\sigma$} &
\colhead {adjusted } &
\colhead {5 $\sigma$} \\
\colhead {} &
\colhead {int times (sec)} &
\colhead {$\mu$ Jy}  &
\colhead {int times (sec)} &
\colhead {$\mu$ Jy}
}
\startdata
5.6	&	 666  &	0.38  & 655 & 0.21	\\
7.7	&	 443  & 0.73	& 866 &	0.20 \\
10	&	 335 & 1.41	& 644 &	0.39 \\
12.8	&	 223  & 2.8	& 755 &	0.62 \\
15	&	 335 & 3.0 & 1121 &	0.75 \\
18	&	 443 & 5.0	& 755 & 1.8	\\
21	&	 2999 & 3.3	& 2187 &	2.8 \\
25.5	&	 623  & 25	& 833 & 17 \\
\enddata
\tablenotetext{a}{The original detection levels are nominal, based on pre-launch information. The detection levels for the adjusted integration times reflect the realized survey performance. They represent improvements by factors 1.7 - 2.6 over prelaunch predictions out through F1800W.}
\label{tab:inttimes}
\end{deluxetable}

\section{Acknowledgements}

We thank the MIRI instrument team for the dedication in building and testing the instrument, as described by \citet{wright2023}, with team members listed as co-authors. We thank Jane Morrison for assistance in reducing the SMILES data. We received helpful comments on the text from Zhiyuan Ji, Yang Sun, and Yongda Zhu.  Work on this paper was supported in part by grant 80NSSC18K0555, from NASA Goddard Space Flight Center to the University of Arizona. IS acknowledges funding support from the Atracc{\' i}on de Talento program, Grant No. 2022-T1/TIC-20472, of the Comunidad de Madrid, Spain. PGP-G's contributions were funded by grant PID2022-139567NBI00 funded by Spanish Ministerio de Ciencia e Innovaci\'on  MCIN/AEI/10.13039/501100011033, FEDER {\it Una manera de hacer Europa}. The work of CCW is supported by NOIRLab, which is managed by the Association of Universities for Research
in Astronomy (AURA) under a cooperative agreement
with the National Science Foundation. 

{\it Facilities: JWST (MIRI)}

{\it Software: Excel, Photoshop}

\eject

\end{document}